\documentclass[lettersize,journal]{IEEEtran}
\usepackage{amsmath,amsfonts,bm,amssymb}
\usepackage{algorithmic}
\usepackage{algorithm}
\usepackage{array}
\usepackage{booktabs}
\usepackage[caption=false]{subfig}
\usepackage{textcomp}
\usepackage{newtxtext,newtxmath}
\usepackage{times}
\usepackage{stfloats}
\usepackage{url}
\usepackage{verbatim}
\usepackage{graphicx}
\usepackage{cite}
\usepackage[colorlinks, linkcolor=blue, citecolor=blue]{hyperref}
\usepackage{enumerate}
\usepackage{multirow}
\usepackage{color}
\usepackage{tabularray}
\usepackage{tabularx}
\usepackage{xcolor}
\usepackage{capt-of}
\usepackage{xcolor}
\begin{document}

\title{Reflection-Driven Self-Optimization 6G Agentic AI RAN via Simulation-in-the-Loop Workflows}

\author{Yunhao Hu, Xinchen Lyu, Chenshan Ren, Keda Chen, Qimei Cui, and Xiaofeng Tao
\thanks{Y. Hu, X. Lyu, K. Chen, Q. Cui, and X. Tao are with  National Engineering Research Center for Mobile Networking Technologies, Beijing University of Posts and Telecommunications, China.
C. Ren is with the Key Laboratory of Ethnic Language Intelligent Analysis and Security Governance of MOE, Minzu University of China, China.}
}

\markboth{Journal of \LaTeX\ Class Files,~Vol.~14, No.~8, August~2021}%
{Shell \MakeLowercase{\textit{et al.}}: A Sample Article Using IEEEtran.cls for IEEE Journals}

\maketitle

\begin{abstract}
The escalating complexity of sixth-generation (6G) networks demands unprecedented levels of autonomy beyond the capabilities of traditional optimization-based and current AI-based resource management approaches. While agentic AI has emerged as a promising paradigm for autonomous RAN, existing frameworks remain fundamentally open-loop, excelling at reasoning and planning but lacking empirical validation mechanisms to verify decisions and learn from outcomes. This article identifies simulation-in-the-loop validation as a critical enabler for autonomous networks, where AI agents can empirically verify decisions and learn from outcomes. We present the first reflection-driven self-optimization framework that integrates agentic AI with high-fidelity network simulation in a closed-loop architecture. Our system orchestrates four specialized agents, including scenario, solver, simulation, and reflector agents, working in concert to transform unverified open-loop planning into a continuous, evidence-grounded, self-optimization workflow. Extensive experiments validate significant performance improvements over non-agentic baselines: 17.1\% higher throughput in interference optimization, 67\% improved user QoS satisfaction through intent recognition, and 25\% reduced resource utilization during low-traffic periods while maintaining service quality.
\end{abstract}

\section{Introduction}
\IEEEPARstart{T}{he} integration of artificial intelligence (AI) and communication has been recognized by the International Telecommunication Union (ITU-T) as a key scenario for the sixth-generation (6G) wireless networks. This vision is driving a fundamental transformation in radio access network (RAN) design, where traditional optimization-based resource management is increasingly being augmented (in some cases replaced) by AI-enabled methods. This paradigm shift accelerates the adoption of data-driven AI techniques in 5G-Advanced and early 6G research, offering promising gains in handling non-convex problems and unseen network scenarios~\cite{cui2025overview}.

However, both traditional optimization and current AI-based approaches for resource management may not achieve the cognitive autonomy essential for 6G RAN. These approaches typically function within fixed system models and optimization objectives, lacking the ability to comprehend high-level user intent, dynamically reformulate problems, or learn from experience without human intervention. This gap is being addressed by the emerging paradigm of agentic AI RAN~\cite{xiao2025towards}, which envisions networks governed by autonomous agents capable of perception, reasoning, and strategic action. Gaining significant traction in both academia and standardization (including 3GPP, ETSI, and 6GANA), agentic AI RAN promises transformative benefits, including intent-driven operation, dynamic adaptation to novel scenarios, and a substantial reduction in operational expenditure (OpEx) by minimizing manual engineering.

\begin{figure*}[!t]
\centering
\subfloat[Tier 1: Direct LLM Use]{
	\includegraphics[width=0.31\textwidth]{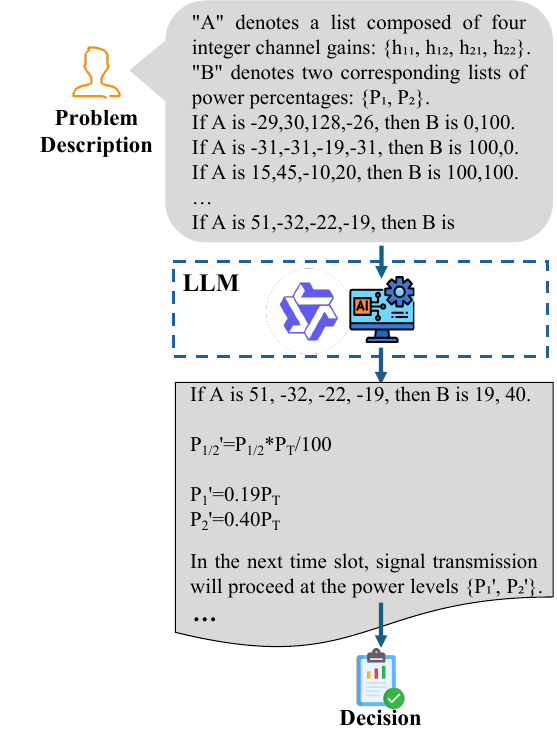}
	\label{1-a}
	}
\hfil
\subfloat[Tier 2: LLM as Tools]{
		\includegraphics[width=0.31\textwidth]{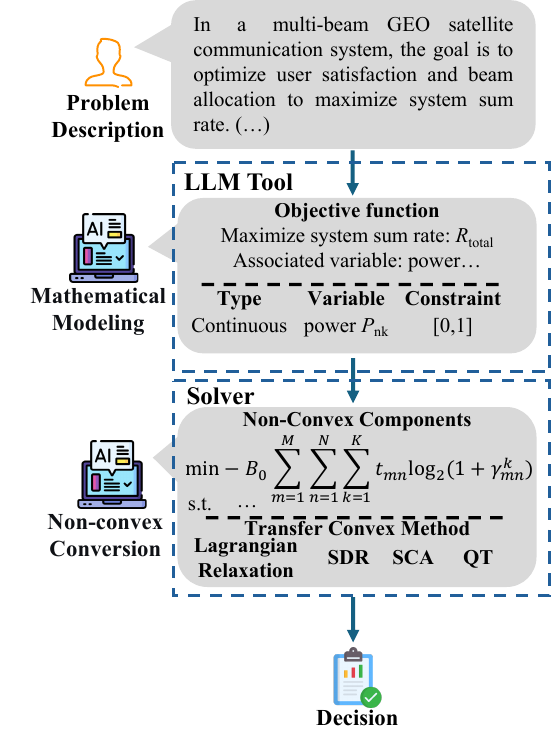}
		\label{1-b}
		}
\hfil
\subfloat[Tier 3: Agentic Workflow]{
		\includegraphics[width=0.31\textwidth]{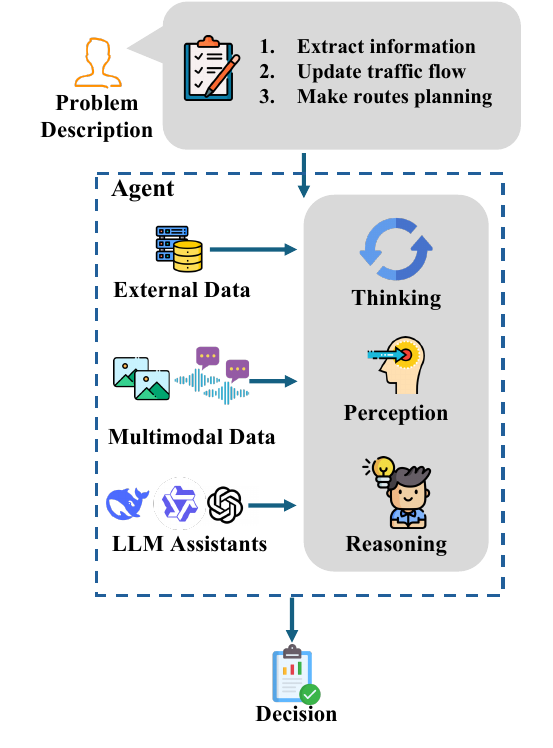}
		\label{1-c}
		}
\quad
\subfloat[Proposed Reflection-Driven Self-Optimization Agentic Framework via Simulation-in-the-Loop Workflows]{
		\includegraphics[width=\textwidth]{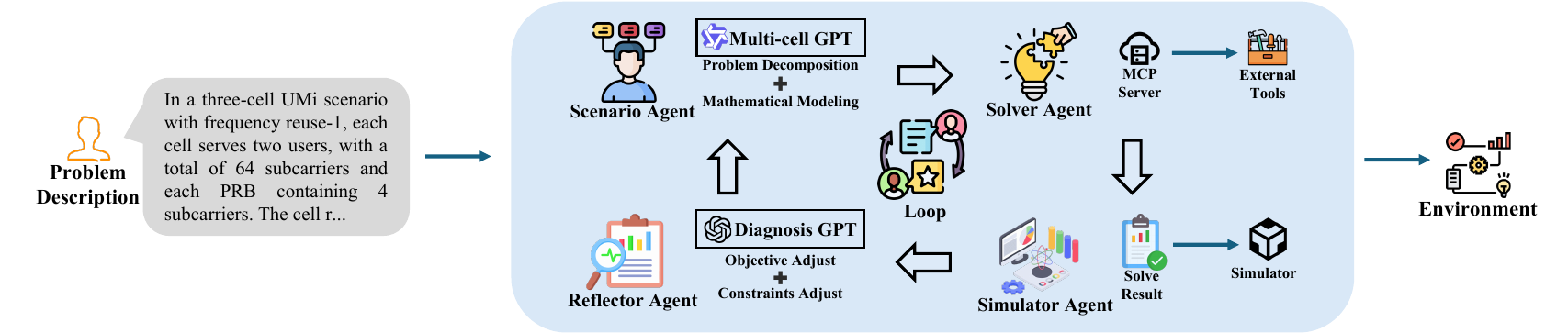}
		\label{1-d}
		}
\caption{Illustrative examples of existing agentic RAN paradigms and the proposed closed-loop architecture. (a)–(c) Current workflows (LLM-direct, LLM-as-tools, and agentic planning) operate as open-loop systems that lack empirical validation. (d) Proposed reflection-driven framework that integrates simulation-in-the-loop validation to enable evidence-grounded, closed-loop self-optimization.}
\label{fig_1}
\end{figure*}

As illustrated in Fig. \ref{fig_1} (a)-(c), recent explorations in agentic AI RAN can be broadly categorized into three types: \textit{(1) Large Language Model (LLM) Direct for Resource Management~\cite{lee2024llm,noh2025adaptive,zhou2025prompting}:} This approach employs the reasoning chain of LLMs to directly output resource allocation decisions. While simple to implement, it often proves inefficient and lacks the precision required for complex, high-dimensional optimization in dense RAN scenarios. \textit{(2) LLM as Tools for Resource Management~\cite{habib2025llm,peng2025llm,he2024designing}:} The LLM acts as analyzer and code generator, interpreting network problems and producing snippets for traditional solvers. While it can achieve similar performance with conventional optimization approaches, its effectiveness is heavily reliant on extensive task-specific fine-tuning and meticulous prompt engineering, limiting its generality. \textit{(3) Agentic Workflow for Resource Management~\cite{elkael2025agentran,xu2024large,tong2025wirelessagent,zhang2025toward,salama2025edge,zhao2025agentification,pellejero2025agentic}:} This method exploits the advanced AI agent with a structured workflow for problem thinking, perception, and multi-step reasoning, which is more applicable for complicated resource optimization tasks.

Despite the sophisticated reasoning capabilities, current agentic AI RAN frameworks ~\cite{elkael2025agentran,xu2024large,tong2025wirelessagent,zhang2025toward,salama2025edge,zhao2025agentification,pellejero2025agentic} operate as fundamentally open-loop systems, lacking mechanisms for grounded reflection and empirical validation. \textit{Without the ability to test hypotheses and learn from outcomes, the reasoning capability of agent remains an unverified, open-loop process, limiting both its performance and generalizability across diverse network conditions.} This necessitates integrating a reflection capability, where the agent can critically analyze its actions through RAN simulation, creating a closed-loop system for continuous self-improvement: (1) Such simulation-in-the-loop design enables the system to move beyond static problem-solving to dynamic, evidence-based adaptation for handling the unpredictable complexity of 6G RAN. (2) The simulation-in-the-loop validation is bridging the gap between autonomous reasoning and reliable, high performance operation.

This article proposes a novel reflection-driven self-optimization framework for 6G agentic AI RAN via simulation-in-the-loop workflows. As illustrated in Fig. \ref{fig_1}\subref{1-d}, our architecture consists of four specialized agents: (1) Scenario Agent that interprets network conditions and decomposes problems; (2) Solver Agent that applies appropriate mathematical techniques to generate resource allocation strategies; (3) Simulation Agent that creates high-fidelity RAN simulation to evaluate proposed solutions; and (4) Reflector Agent that analyzes performance outcomes and guides iterative improvements. Our key contributions are as follows:\par

\begin{itemize}
  \item We propose the first comprehensive reflection-driven framework that integrates agentic AI with RAN simulation in a closed-loop workflow, enabling autonomous validation and refinement of optimization strategies.
  \item We formalize four specialized agents with clearly defined interfaces and responsibilities that collectively enable closed-loop autonomous network optimization.
  \item We design three use cases to show the performance gains of proposed agentic RAN architecture, where the simulation-in-the-loop design allows the system to escape local optima, recognize implicit user intent, and dynamically adapt to changing network contexts.
\end{itemize}

Extensive simulations demonstrate substantial gains over conventional non-agentic baselines, including a 17.1\% throughput increase in interference management, a 67\% improvement in user QoS satisfaction through intent-aware allocation, and a 25\% reduction in resource utilization during low-traffic periods while maintaining strict service guarantees.

\section{Preliminaries: Evolution Towards Agentic AI RAN}
This section briefly reviews the technological evolution toward agentic AI RAN, including the efforts for AI-RAN integration and recent LLM/agentic techniques, and discusses the future reflection-driven agentic RAN.

\subsection{AI-RAN Integration in Standardization Landscape}

Major standardization bodies have established critical frameworks for AI integration into RAN architectures. The O-RAN Alliance has defined the RAN Intelligent Controller (RIC) architecture with A1/E2 interfaces enabling AI-driven control loops, while 3GPP has initiated research items for AI/ML management (TS 28.104) and the Network Data Analytics Function (NWDAF, TS 23.288). These standards transform RAN operations through three key shifts:

(1)~\textit{Prediction replaces modeling}: AI/ML models for channel state prediction and traffic forecasting are now specified components for proactive resource allocation.

(2)~\textit{Data-driven control augments algorithmic solvers}: The RIC/xApp framework enables ML-based policies (e.g., for mobility management) to operate alongside or in place of traditional optimization solvers.

(3)~\textit{AI automation supersedes manual configuration}: Standards now explicitly define ML model training and inference workflows that automate tasks previously requiring expert human intervention.

This evolution represents a fundamental transition from purely model-based optimization to data-driven, AI-augmented RAN operations, establishing the technical foundation upon which agentic systems can build.

\begin{figure*}[!htbp]
\centering
	\label{tab:literature_review}
    \renewcommand{\arraystretch}{1.5}
    \footnotesize 
	\begin{tabularx}{\textwidth}{| 
        >{\centering\arraybackslash\bfseries}m{2.5cm} | 
        >{\centering\arraybackslash}m{1.5cm} | 
        >{\raggedright\arraybackslash}X |}
        
        \hline
        Research Direction & \textbf{Literature} & \centering \textbf{Highlight} \arraybackslash \\ 
        \hline
        
        \multirow{3}{=}{\centering Tier 1. LLMs For Direct Problem Solving} 
        & \cite{lee2024llm} & Leverages LLM knowledge transfer to maximize energy and spectrum efficiency in resource allocation. \\ \cline{2-3}
        
        & \cite{noh2025adaptive} & Uses prompt-based tuning to convey dynamic QoS constraints for constraint-aware resource allocation. \\ \cline{2-3}
        
        & \cite{zhou2025prompting} & Achieves optimization without model fine-tuning via reinforced in-context learning for power control. \\ \hline
        
        \multirow{3}{=}{\centering Tier 2. LLMs As Tools For Problem Solving} 
        & \cite{habib2025llm} & Extracts high-level intents via LLMs and optimizes 5G/O-RAN using hierarchical reinforcement learning. \\ \cline{2-3}
        
        & \cite{peng2025llm} & Detects and converts non-convex components into solvable forms for automated resource allocation solutions. \\ \cline{2-3}
        
        & \cite{he2024designing} & Leverages LLM generative capabilities to autonomously design diverse network optimization algorithms. \\ \hline
        
        \multirow{8}{=}{\centering Tier 3. Agentic Workflow For Problem Solving} 
        & \cite{elkael2025agentran} & Generates adaptive control algorithms from natural language intents for hierarchical autonomous 6G networks. \\ \cline{2-3}
        
        & \cite{xu2024large} & Integrates diverse tools for cross-domain operations like topology discovery and physics-informed optimization. \\ \cline{2-3}
        
        & \cite{tong2025wirelessagent} & Constructs an autonomous agent framework with perception and planning for adaptive network slice management. \\ \cline{2-3}
        
        & \cite{zhang2025toward} & Enables edge devices to evolve into autonomous agents via a collaborative perception-action closed loop. \\ \cline{2-3}
        
        & \cite{salama2025edge} & Achieves real-time, multi-level autonomous optimization in Open RAN environments via edge Agentic AI. \\ \cline{2-3}
        
        & \cite{zhao2025agentification} & Enables continuous self-adaptation and evolution from fixed to mobile antenna optimization tasks. \\ \cline{2-3}
        
        & \cite{pellejero2025agentic} & Combines time series analysis with agents for intent decomposition and anomaly-based strategy execution. \\ \hline
        
    \end{tabularx}
  \par
  \vspace{0.1cm}
  \includegraphics[width=0.8\textwidth]{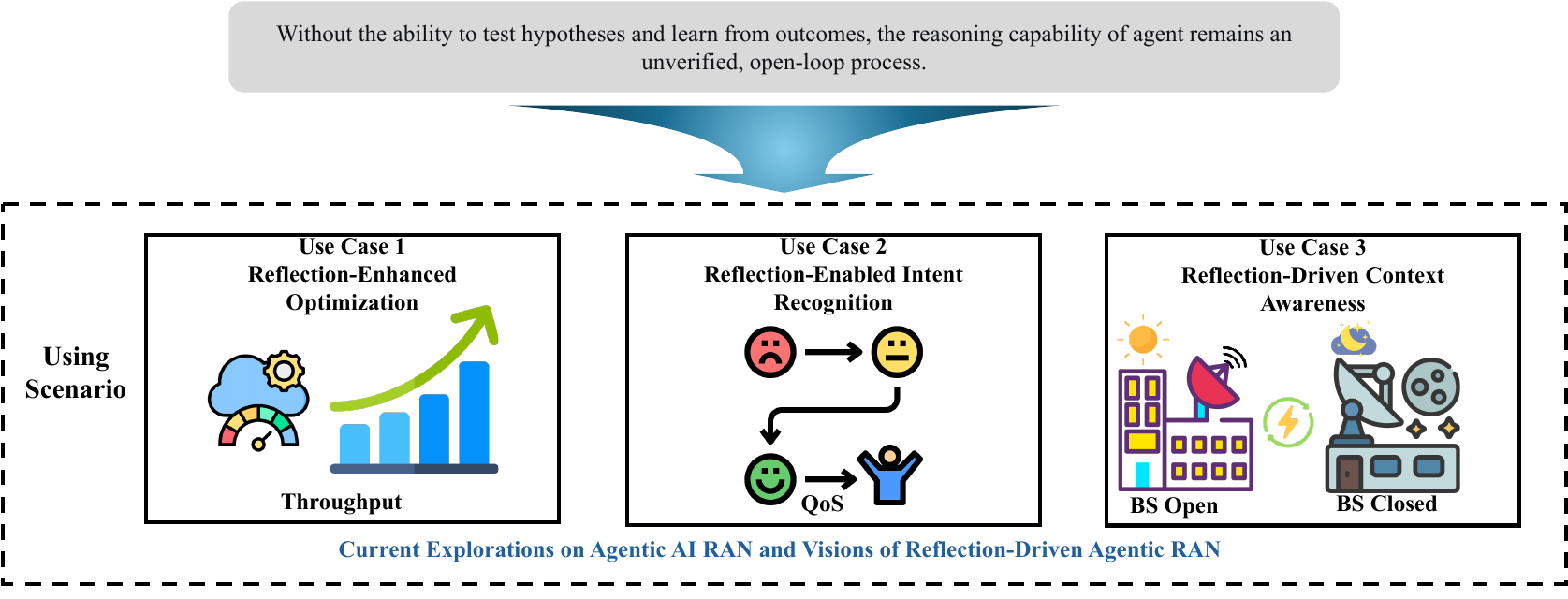}
  \captionof{figure}{Difference between existing agentic AI RAN paradigms and the proposed closed-loop vision. Different from existing open-loop approaches, the proposed framework establishes a simulation-in-the-loop closed-loop workflow, thereby transforming unverified AI reasoning into evidence-grounded, self-optimization network autonomy.}
  \label{fig2}
\end{figure*}

\subsection{Evolution towards LLM and Agentic AI RAN}

As shown in Fig. \ref{fig2}, recent research explores LLMs and agentic AI techniques as catalysts for autonomy, evolving through three distinct tiers.

\subsubsection{Tier 1: LLM Direct for Resource Management}
This approach employs the inherent reasoning capability of LLM to directly map network states to resource allocation decisions through prompt engineering. In \cite{lee2024llm}, the LLM leverages its pre-trained knowledge base to maximize energy or spectrum efficiency. In~\cite{noh2025adaptive}, specialized prompting strategies are developed to explicitly convey QoS constraints within the natural language prompt, enabling constraint-aware solutions. In~\cite{zhou2025prompting}, reinforced in-context learning is applied to achieve power control optimization without model fine-tuning. However, these methods lack the precision of mathematical solvers, proving inefficient for high-dimensional continuous optimization problems like precise power control in dense networks, where gradient-based methods outperform LLM reasoning.

\subsubsection{Tier 2: LLM as Tools for Resource Management} This paradigm demonstrates a shift from direct LLM solution to using LLMs as specialized tools to generate or augment traditional optimization pipelines. In~\cite{habib2025llm}, LLMs are used to extract operational intent from natural language and interface with reinforcement learning modules for actual optimization. In~\cite{peng2025llm}, LLMs are designed to identify and reformulate problematic components before passing them to conventional solvers. In~\cite{he2024designing}, the generative capabilities of LLMs are exploited to produce diverse algorithm design schemes. While achieving performance comparable to human-designed optimizers, these approaches require task-specific fine-tuning and carefully crafted prompts for each new problem type.

\subsubsection{Tier 3: Agentic Workflow for Resource Management} This paradigm embeds the LLM to construct an agentic workflow of tool composition and multi-step planning. In~\cite{elkael2025agentran}, a hierarchical agent system is designed to translate natural language intents into continuously improved control algorithms. In~\cite{xu2024large}, LLM-driven systems integrate diverse tools for complex cross-domain operations spanning real-time topology discovery and physical network reconfiguration. In~\cite{tong2025wirelessagent}, an autonomous agent framework equipped with advanced perception and planning capabilities is constructed for adaptive network slice management. In~\cite{zhang2025toward} and \cite{salama2025edge}, perception-reasoning-action loops transform passive edge devices into collaborative agents with independent cognition. In~\cite{zhao2025agentification}, continuous self-adaptation is enabled to evolve from fixed to mobile antenna optimization tasks. In~\cite{pellejero2025agentic}, time series analysis is combined with agents for intent decomposition and anomaly-based strategy execution. 

\textit{The Open-Loop Limitation.} Despite their increasingly sophisticated reasoning capabilities, all three tiers~\cite{elkael2025agentran,xu2024large,tong2025wirelessagent,zhang2025toward,salama2025edge,zhao2025agentification,pellejero2025agentic} operate as fundamentally open-loop systems. Prior workflows excel at reasoning and action generation but lack mechanisms to empirically validate decisions against realistic network feedback or learn from execution outcomes. Consequently, the agent's reasoning remains an unverified process, severely limiting its ability to diagnose whether performance degradation stems from misaligned objectives, incomplete constraints, or even flawed problem decomposition. This critical gap necessitates a paradigm shift toward a closed-loop agentic architecture that integrates continuous, simulation-grounded validation.

\subsection{The Vision for Reflection-Driven Closed-Loop Agents}

To transcend the limitations of open-loop planning, we advocate for a simulation-in-the-loop architecture that transforms agentic RAN from a sophisticated planner into a self-correcting optimizer. By embedding a high-fidelity digital twin within the agent workflow, the system establishes a closed validation loop where each candidate strategy is empirically tested, evaluated against quantitative KPIs, and iteratively refined through LLM-driven reflection. 

As shown in the bottom of Fig. \ref{fig2}, reflection-driven self-optimized agentic framework improves RAN performance in the following aspects: (1)~\textit{Reflection-Enhanced Optimization (Use Case 1)}: Closed-loop reflection empowers the agent to analyze simulation feedback, recognize optimization plateaus, and autonomously redirect search paths by adjusting constraints or objectives, thereby discovering superior global solutions inaccessible to conventional methods. (2)~\textit{Reflection-Enabled Intent Recognition (Use Case 2)}: By reflecting on QoE patterns, the agent infers implicit intent from performance disparities and proactively rebalances resources to align with actual service expectations. (3)~\textit{Reflection-Driven Context Awareness (Use Case 3)}: The closed-loop design allows the agent to monitor environmental shifts through simulation outcomes and autonomously transition between operational priorities (e.g., from spectral efficiency during peak hours to energy conservation during off-peak periods) without manual reconfiguration.

\section{Proposed Reflection-Driven Self-Optimization 6G Agentic AI RAN Framework}
This section presents our comprehensive framework for enabling autonomous self-optimization in 6G RAN through reflection-driven methodologies. We begin by establishing the fundamental design rationale for reflection and simulation-in-the-loop workflow, present the framework architecture, and discuss the key design and implementation details. 

\subsection{The Rationale of Reflection and Simulation-in-the-Loop}

\subsubsection{Reflection is the cornerstone of our framework} Reflection empowers the system to achieve autonomous self-optimization in 6G RAN. In dynamic RAN environments characterized by user mobility, divergent quality-of-service (QoS) requirements, and time-varying traffic patterns, traditional optimization methods often converge to suboptimal solutions due to rigid problem formulations or myopic decision-making. Reflection addresses this by allowing the agent to iteratively assess simulation outcomes, identify limitations in current optimization strategies (e.g., local optima in integer programming for resource block allocation), and adapt its approach to evolving network conditions. \par For instance, when optimizing inter-cell interference (ICI) in multi-cell scenarios, reflection enables the agent to dynamically reevaluate whether prioritizing spectral efficiency over energy savings aligns with current user demands or network congestion levels. This capability is critical for RANs, where static optimization frameworks fail to account for real-world uncertainties such as sudden traffic spikes or device mobility, thereby ensuring sustained performance under heterogeneous and unpredictable operating conditions.

\subsubsection{Simulation-in-the-loop serves as the foundational enabler of reliable reflection-driven optimization} By embedding a high-fidelity RAN simulator within the workflow, simulation-in-the-loop provides a controlled environment for rigorously testing optimization decisions before deployment, mitigating risks of system instability caused by erroneous configurations. Unlike offline optimization, which relies on simplified analytical models, simulation-in-the-loop leverages system-level simulations to generate realistic KPIs such as throughput, SINR, and power consumption under complex channel conditions. This validates the feasibility of proposed solutions and enhances the trustworthiness of the agentic process by exposing the agent to edge cases (e.g., high-mobility user scenarios) that analytical models may overlook. Crucially, simulation-in-the-loop ensures that reflection is grounded in empirical evidence rather than theoretical assumptions, transforming the agent from a ``black-box'' optimizer into a self-correcting system capable of learning from simulated failures without disrupting live networks.

\subsection{Framework Architecture}

\begin{figure*}[!htbp]
  \centering
  \includegraphics[width=0.95\textwidth]{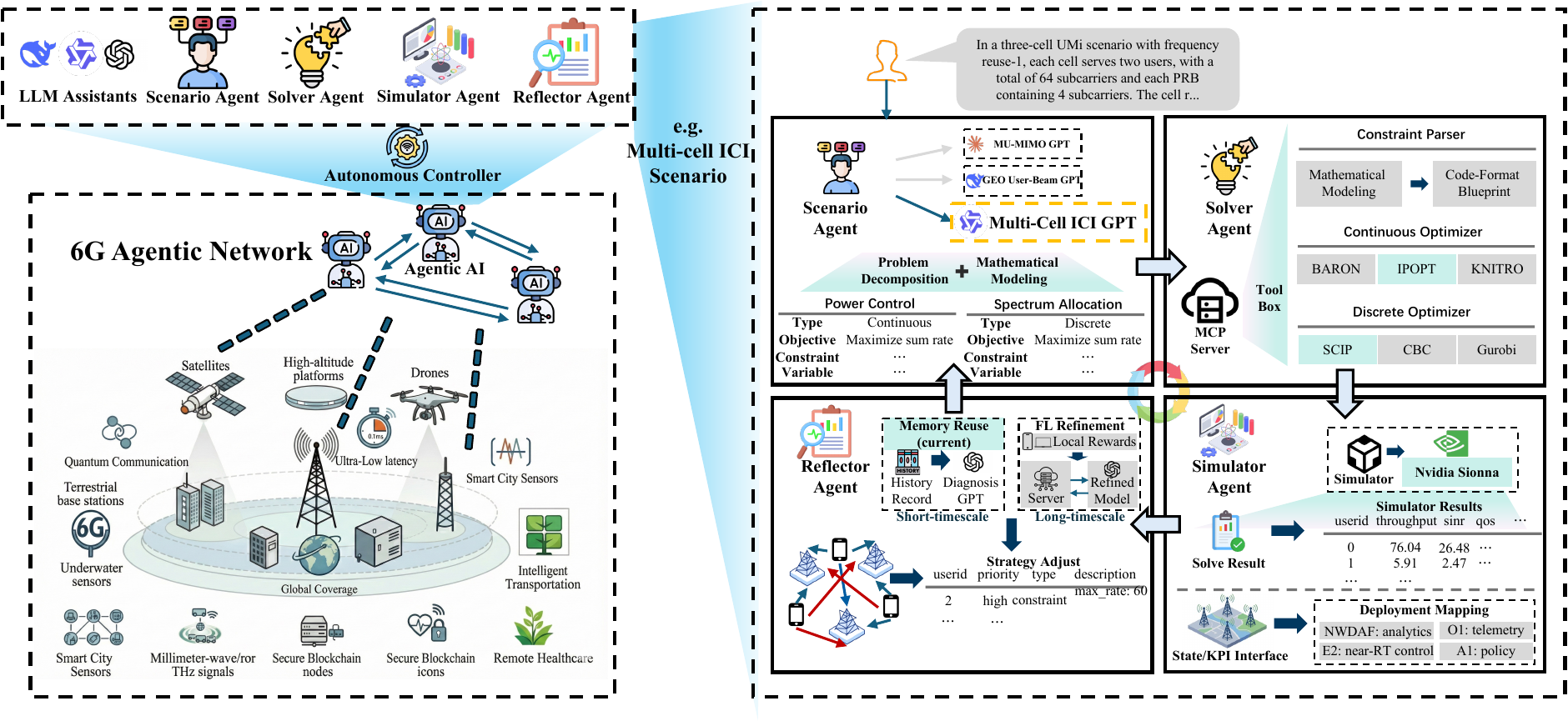}
  \caption{Architecture of the proposed reflection-driven self-optimization framework. The closed-loop workflow orchestrates four specialized agents (i.e., Scenario, Solver, Simulation, and Reflector) linked by structured interfaces and standardized external mappings.}
  \label{fig3}
\end{figure*}
As shown in Fig. \ref{fig3}, the proposed framework comprises four specialized agents, each addressing distinct stages of the optimization workflow while ensuring seamless interoperability. 

\subsubsection{Scenario Agent} Acting as the initial interpreter, this agent translates high-level user directives  into structured network configurations and tractable subproblems. By leveraging retrieval-augmented generation (RAG) grounded in domain-specific literature, it decomposes complex optimization tasks into interdependent components (such as joint resource block allocation and power control) ensuring mathematical rigor and scalability across large-scale topologies.

\subsubsection{Solver Agent} Tasked with mathematical execution, this agent selects appropriate optimization algorithms (e.g., IPOPT for continuous convex problems, CBC for integer programming) based on the structural decomposition provided by the Scenario Agent. It interfaces with the simulation environment to acquire real-time network states (e.g., channel state information) and converts solver outputs into actionable configuration vectors in the simulation layer.

\subsubsection{Simulation Agent} Serving as the sandbox for verification, this agent is built upon the GPU-accelerated Sionna platform to execute high-fidelity, system-level RAN emulations. It instantiates the network topology defined by the Scenario Agent, applies the resource allocation strategies generated by the Solver Agent, and emulates network dynamics over a specified timeframe. By parallelizing physical-layer computations, the agent reduces evaluation latency from hours to minutes, enabling rapid iterative refinement. Upon completion, it extracts comprehensive KPIs, and packages them for reflection.

\subsubsection{Reflector Agent} Equipped with advanced reasoning capabilities, this agent performs the critical diagnostic function of the closed loop. Through a structured ReAct process, it analyzes simulation outputs to identify performance bottlenecks, constraint violations, or suboptimal trade-offs. It autonomously formulates strategic refinements (such as adjusting QoS priorities, reformulating objectives from sum-rate to proportional fairness, or tightening resource bounds) and feeds these directives back to the Scenario Agent. This feedback mechanism effectively seals the autonomous self-optimization cycle.

While the current framework derives state observations and KPIs exclusively from a digital twin, the architecture's standardized abstraction layer is explicitly engineered for seamless extension to physical-world interfaces. This unified data schema can be directly mapped to industry-standard open APIs (such as 3GPP NWDAF for predictive analytics and O-RAN's O1, E2, and A1 interfaces for telemetry, near-real-time control, and policy enforcement), enabling the closed-loop reflection mechanism to transition from simulated validation to live, production-grade autonomous 6G RAN deployments.

\subsection{Key Design Elements}

The practical realization of this framework involves addressing several key design challenges.

\subsubsection{Simulator Integration and Bounded Action Space}

Integrating a high-fidelity RAN simulator into a dynamic, iterative agentic loop is operationally complex, demanding both rigorous interface control and computational efficiency. Our framework establishes a standardized API and unified data schema that govern all state, configuration, and KPI exchanges. Crucially, the agentic reasoning layer does not emit raw simulator commands or free-form updates. Instead, the Solver and Reflector Agents operate within a strictly bounded action space of typed, localized modifications, such as constraint tightening, per-user/per-cell cap tuning, and stage-level strategy switching. A dedicated simulator bridge translates these high-level directives into structured requests containing validated scenario parameters, topology configurations, traffic context, and KPI bounds before invoking the simulation engine. Every request and returned KPI bundle undergoes comprehensive validation for field completeness, numerical consistency, and cross-record alignment (user-, cell-, and KPI-level), ensuring that reflection outputs interact with the simulator only through a machine-checkable interface.

\subsubsection{Domain-Grounded Problem Decomposition}

General-purpose LLMs risk misinterpreting complex wireless scenarios, which would propagate structural errors throughout the closed-loop workflow. To ensure technical accuracy, we equip the Scenario Agent with a specialized retrieval-augmented generation (RAG) system grounded in a curated database of several optimization literature. By retrieving contextually relevant formulations and solution methodologies before decomposition, the agent reliably maps high-level intents to mathematically sound subproblems (e.g., correctly separating joint user association and power control into a tractable two-stage process). This domain-grounded parsing minimizes decomposition-induced failures and ensures that subsequent solver and simulation phases operate on structurally valid problem representations.

\subsubsection{Phased Human-in-the-Loop Calibration}

To bootstrap reflection reliability in novel problem classes, we implement a structured Human-in-the-Loop (HITL) initialization mechanism. During the first 5–10 optimization cycles for a new scenario, the Reflector Agent generates multiple candidate refinement strategies, and a domain expert selects the most viable option. This guided calibration accelerates convergence to stable KPI trajectories and aligns the reflection logic with domain-specific trade-offs. Once consistent performance gains and reliable diagnostic behavior are observed, mandatory HITL is phased out, transitioning the system to fully autonomous closed-loop operation. Crucially, HITL remains available as an escalation safeguard for low-confidence decisions, anomalous network states, or iterations that trigger rollback conditions.

\begin{figure*}[!htbp]
  \centering

  \includegraphics[width=0.9\textwidth]{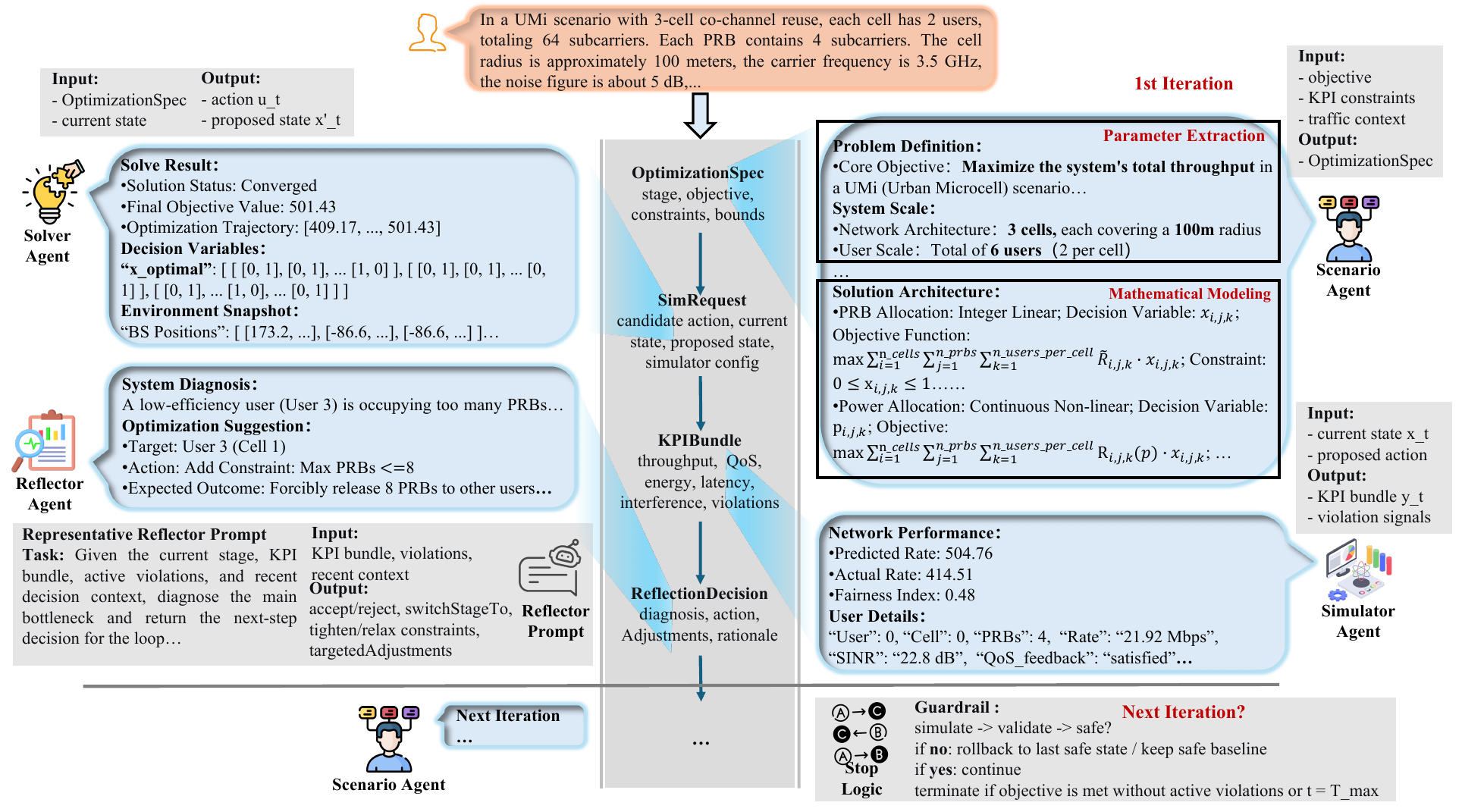}
  \caption{Closed-loop agentic workflow for reflection-enhanced optimization (Use Case 1). The diagram illustrates the iterative cycle across four standardized interfaces. The representative trace shows how LLM-driven reflection diagnoses resource imbalances, issues localized constraint adjustments, and converges to a superior solution.}

\label{fig4}
\end{figure*}

\subsubsection{Memory-Driven Adaptation}

Rather than relying on continuous parameter updates, the framework achieves sustained improvement through structured experiential memory in agentic architecture. Each validated iteration (comprising scenario context, executed actions, KPI outcomes, reflection diagnostics, and successful refinements) is archived in a persistent knowledge base. When analogous topologies, traffic patterns, or interference conditions recur, the system retrieves and applies these evidence-backed trajectories, dramatically reducing redundant trial-and-error and accelerating adaptation across heterogeneous deployment scenarios. This memory-driven approach ensures that the closed-loop compounds its intelligence over time, transforming isolated simulation runs into cumulative operational expertise for future optimization decisions.

\subsubsection{Operational Guardrails and Rollback Mechanisms}

Autonomous reflection loops must inherently prevent performance degradation or unstable state propagation. Our architecture enforces a strict separation of proposal, execution, validation, and reflection phases. If a candidate action degrades KPIs, violates constraint thresholds, or underperforms the established safe baseline, the workflow automatically rolls back to the last validated state rather than accepting the update. Every iteration logs structured audit trails (e.g., candidate evaluations, violation signals, reflector rationales, and rollback triggers), ensuring full transparency and inspectability. While these foundational safeguards ensure operational stability, systematic validation under adversarial conditions, noisy KPI streams, and formal safety verification remain critical avenues for future research.

\subsection{Example of Agentic Workflow}

Fig. \ref{fig4} illustrates a representative execution trace of the closed-loop workflow, demonstrating how the four specialized agents interact to transform a static optimization request into a dynamic, self-correcting process. The system is linked by standardized, machine-checkable interfaces that ensure consistent data flow and empirical grounding at every iteration.
(1) Input \& Decomposition. An OptimizationSpec (stage, objective, constraints, context) initializes the cycle. The Scenario Agent parses this into solver-ready subproblems (e.g., separating discrete PRB allocation from continuous power control).
(2) Solve \& Validate. The Solver Agent computes a candidate action (e.g., x\_optimal, p\_optimal) packaged as a SimRequest. The Simulation Agent executes this in Sionna, returning an empirical KPIBundle (throughput, QoS, fairness, violations).
(3) Reflect \& Adapt. The Reflector Agent analyzes the KPIBundle via a structured prompt: ``Given current stage, KPIs, violations, and context, diagnose the bottleneck and return: accept/reject, switchStageTo, tighten/relax constraints, or targetedAdjustments.'' In Fig. 4, it identifies ``User 3 monopolizes PRBs with low efficiency'' and outputs a ReflectionDecision: Target: User 3; Action: Max PRBs $\leq$ 8.
(4) Control Logic. The loop terminates when objectives are met without violations or the iteration budget expires. Unsafe updates trigger automatic rollback; all steps are logged. This evidence-grounded cycle enables autonomous escape from local optima and dynamic adaptation to network feedback.

\section{Case Studies}
This section validates the performance of our reflection-driven framework in a multi-cell RAN for ICI optimization. We evaluate three use cases: (1) Reflection-Enhanced Optimization, where the agent escapes local optima by dynamically redirecting search paths using simulation feedback; (2) Reflection-Enabled Intent Recognition, where the agent infers unmet user QoE requirements from simulated patterns and adapts resource allocation; and (3) Reflection-Driven Context Awareness, where the agent reorients optimization objectives based on environmental changes observed through simulation.

\textbf{Experimental Setup and Benchmarks:} Our evaluation employs a 3-cell Urban Micro (UMi) scenario compliant with 3GPP TR 38.901, with 6 user equipment (UE) devices uniformly distributed within a 100 m cell radius. The physical layer uses OFDM downlink configuration at 3.5 GHz carrier frequency with 180 kHz subcarrier spacing (64 subcarriers, 16 PRBs).\footnote{We also evaluate a larger 7-cell scenario with 21 UEs and 84 PRBs for scalability. The results are provided in the arXiv appendix (\url{http://arxiv.org/abs/2512.20640}).} The proposed framework (where the LLM is Qwen3-Max inference engine) is benchmarked against typical non-agentic resource optimization approach, where the optimizer does not use the LLM for adaptive problem formulation and uses the conventional two-stage solver for joint RB allocation and power optimization. All simulations execute on the Sionna platform with 3GPP channel models, and results are averaged over 100 Monte Carlo trials.

\begin{figure}[!htbp]
\centering
\subfloat[Use Case 1: Reflection-enhanced optimization]{
	\includegraphics[width=3.5in]{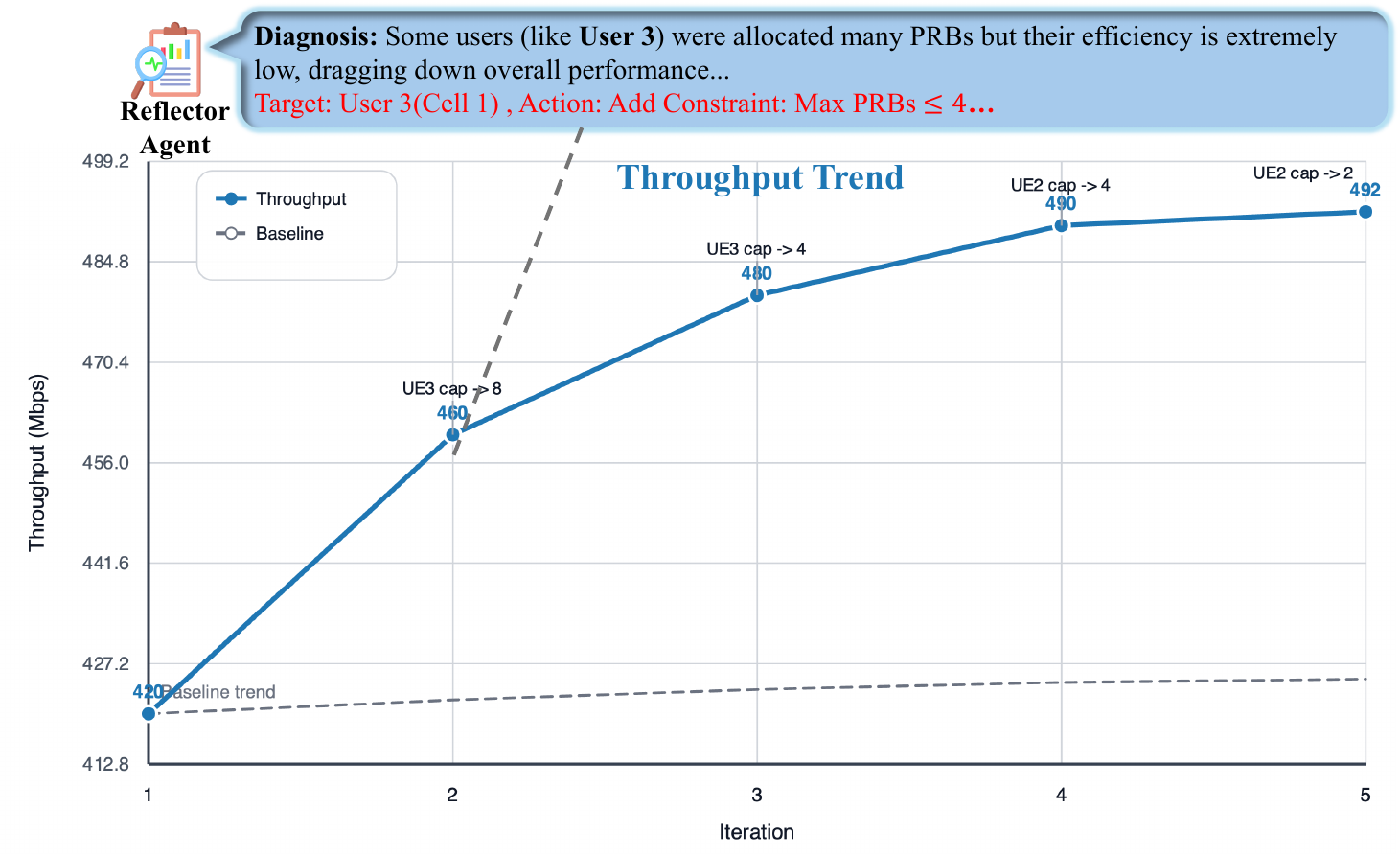}
	\label{5-a}
	}
\vspace{-0.15in}
\subfloat[Use Case 2: Reflection-enabled intent recognition]{
		\includegraphics[width=3.5in]{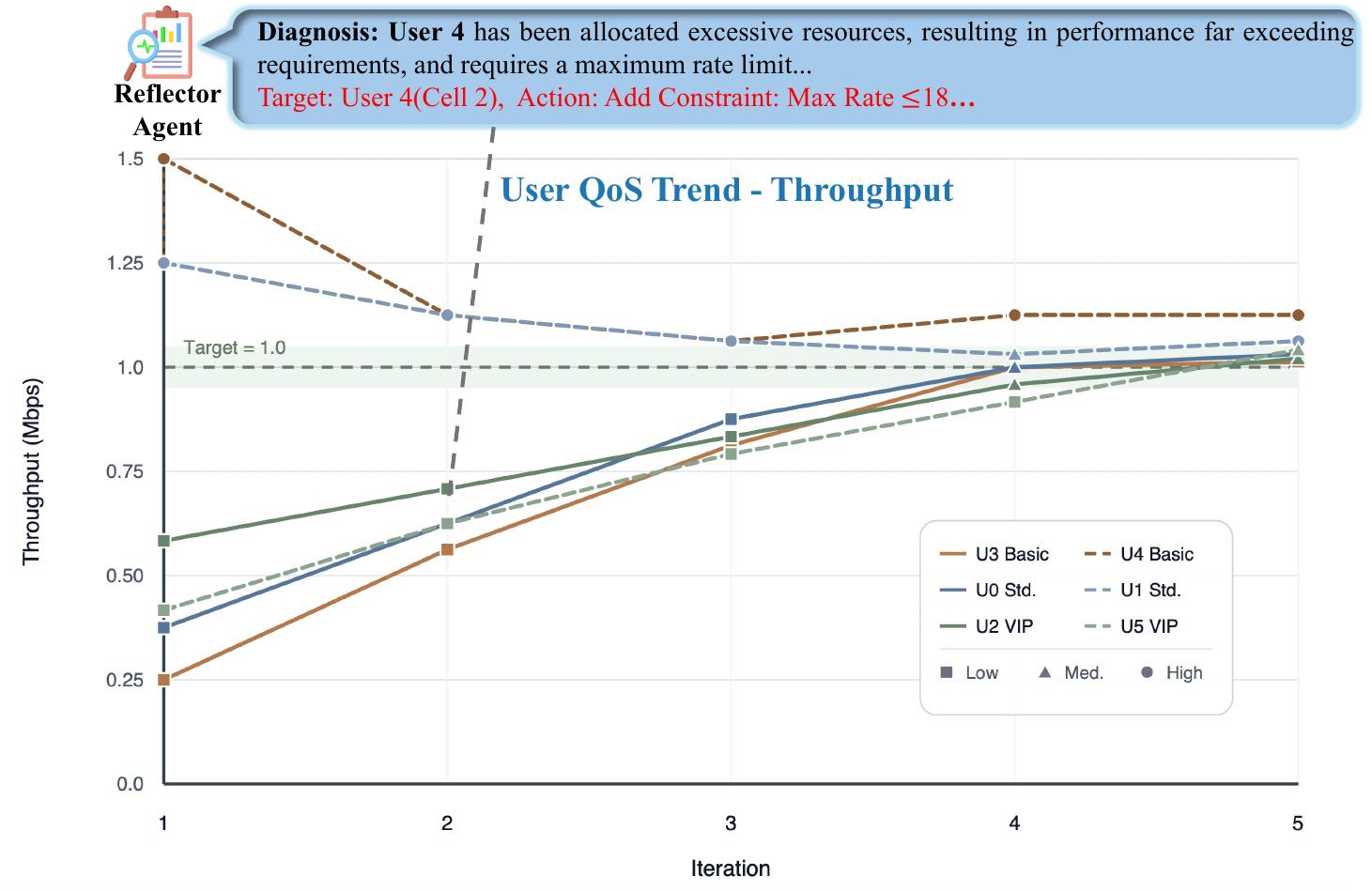}
		\label{5-b}
		}
\vspace{-0.15in}
\subfloat[Use Case 3: Reflection-driven context awareness]{
		\includegraphics[width=3.5in]{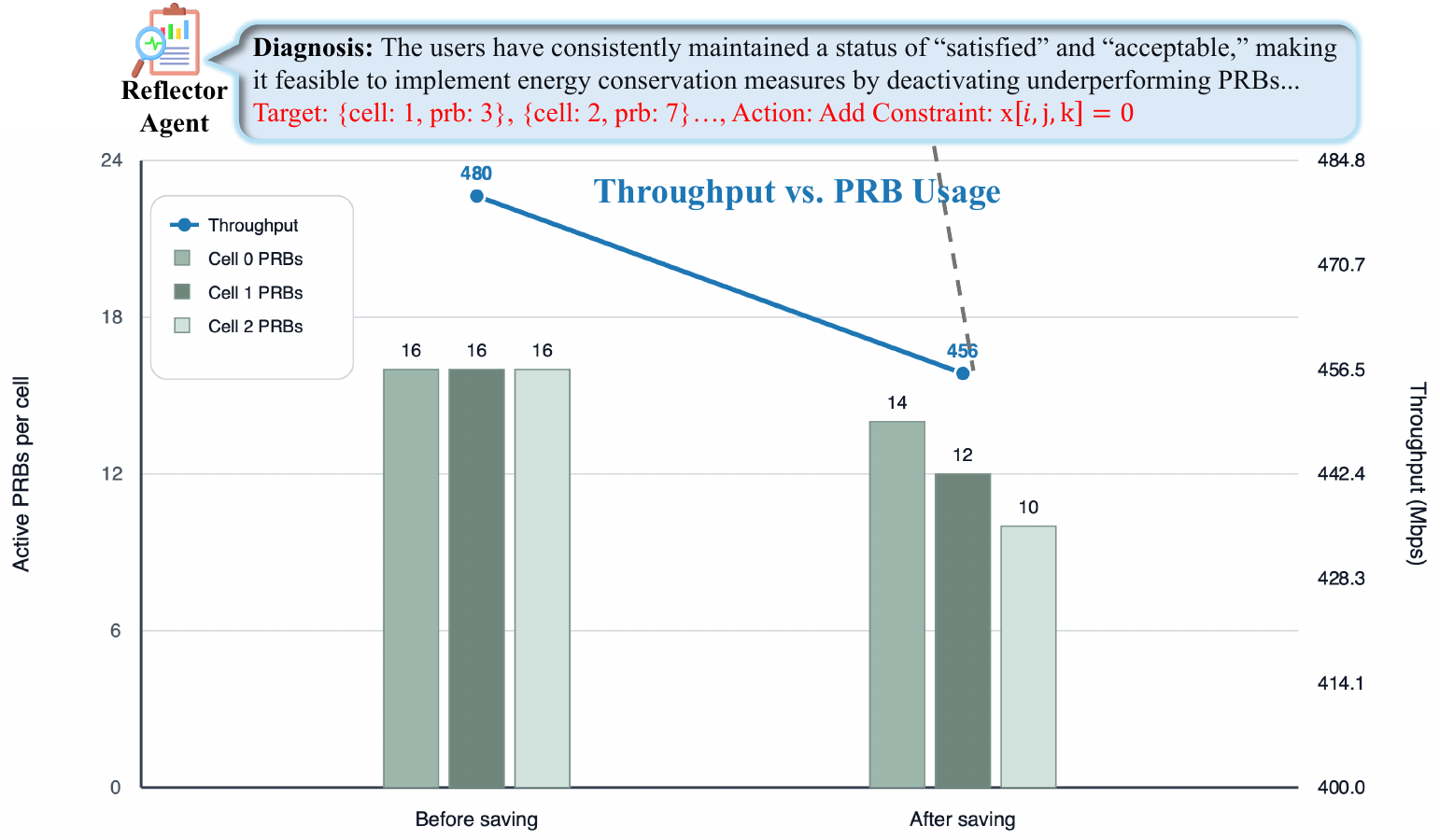}
		\label{5-c}
		}
\caption{Experimental results of proposed reflection-driven agentic framework.}
\label{fig_5}
\vspace{-0.2in}
\end{figure}

\subsection{Case 1: Reflection-Enhanced Optimization}

Fig. \ref{fig_5}\subref{5-a} illustrates the framework's ability to escape local optima in integer-programming-based resource allocation through iterative constraint refinement. The baseline solver converges to a suboptimal throughput of 425 Mbps due to rigid problem formulation and an inability to diagnose resource inefficiencies. In the first iteration, the Reflector Agent analyzes the simulation output and identifies that User 3 (Cell 1) monopolizes PRBs despite exhibiting low spectral efficiency. Acting on this diagnostic feedback, the agent progressively tightens constraints: Iteration 2 caps User 3's allocation at no more than 8 PRBs, boosting throughput to 460 Mbps; Iteration 3 reduces the cap to 4 PRBs while addressing User 2's inefficiency; and Iterations 4 and 5 fine-tune User 2's allocation to no more than 2 PRBs. This progressive refinement culminates in a final throughput of 492 Mbps, corresponding to a 17.1\% gain over the baseline. Unlike open-loop planners that accept solver convergence as final, our framework continuously validates outcomes, diagnoses bottlenecks, and autonomously redirects the search space toward globally superior solutions.

%

\subsection{Case 2: Reflection-Enabled Intent Recognition}

Fig. \ref{fig_5}\subref{5-b} demonstrates how the agent infers implicit user intent from simulated QoS patterns and dynamically adapts resource allocation. Initially, the system maximizes aggregate throughput but yields uneven QoS satisfaction across users. The Reflector Agent detects that User 4 (Cell 2) is over-provisioned while User 3 (Cell 1) remains under-served. Through simulation-grounded reflection, the agent iteratively adjusts objectives: Iteration 2 caps User 4's rate at no more than 18 Mbps; Iteration 3 imposes a minimum rate of at least 6 Mbps for User 3; and Iterations 4 and 5 guarantee baseline rates for Users 0 and 2 while maintaining overall efficiency. This closed-loop adaptation elevates the system to a state where all users achieve satisfactory QoS, increasing the overall QoS satisfaction rate by 67\%. This capability emerges directly from the simulation-in-the-loop validation, which allows the agent to detect intent-performance gaps and autonomously shift its optimization priority from raw throughput maximization to equitable service delivery across users with diverse QoS needs.

\subsection{Case 3: Reflection-Driven Context Awareness}
Fig. \ref{fig_5}\subref{5-c} evaluates the framework's ability to dynamically trade off performance optimization and energy efficiency based on real-time network context. During peak traffic, the agent maintains a performance-first objective, achieving spectral efficiency comparable to traditional optimizers. As network utilization declines, the Reflector Agent identifies that sustaining full-power operation constitutes unnecessary resource waste. It autonomously triggers an energy-saving workflow, selectively deactivating underperforming PRBs while preserving coverage and minimum rate guarantees. As shown in the bar chart, total PRB usage drops from 48 to 36, a 25\% reduction, with only a marginal throughput decrease from 480.0 to 456.0 Mbps, or 5.0\%. The per-cell breakdown (16 to 14 in Cell 0, 16 to 12 in Cell 1, and 16 to 10 in Cell 2) confirms the agent's capacity to dynamically realign optimization objectives with environmental conditions. This context-aware adaptation is only possible through continuous simulation feedback, which grounds operational decisions in empirical network states rather than static, pre-defined rules.

\section{Future Directions for Agentic RAN}
The reflection-driven, simulation-in-the-loop framework establishes a foundational paradigm for autonomous 6G networks. To advance from research prototypes to deployed, trustworthy systems, several critical research directions must be pursued. This section outlines key challenges and opportunities in deployment, coordination, security, and intelligence that will shape the evolution of Agentic RAN.

\subsubsection{Real-Time Digital Twins for Live Deployment}
The transition from simulated environments to live network operation requires overcoming significant technical hurdles through advanced digital twin technology: (1) Real-Time Data Synchronization: Establishing mechanisms for continuous, low-latency data exchange between physical networks and their digital counterparts to maintain accurate state representation. (2) Real-Time Inference and Actuation: Optimizing agent reasoning latency to meet stringent RAN control loop requirements while maintaining decision quality. (3) Safety Guarantees: Developing verification frameworks and constraint enforcement mechanisms to prevent harmful actions and ensure operational safety in live deployments.

\subsubsection{Multi-Agent Collaboration and System Emergence}
Network-wide optimization necessitates sophisticated coordination among specialized agents, opening rich research avenues: (1) Hierarchical Agent Architectures: Designing frameworks for RAN agents to collaborate with higher-layer network agents through standardized interfaces, enabling end-to-end optimization across network domains. (2) Cooperation and Competition Mechanisms: Creating communication protocols and incentive models that balance local objectives with global optimization goals using game-theoretic approaches. (3) Emergent Behavior Management: Understanding and controlling complex system-wide behaviors that arise from multi-agent interactions to ensure network stability and performance.

\subsubsection{Robustness and Security of Autonomous Decisions}
The autonomous nature of Agentic RAN introduces unique security challenges that demand proactive solutions: (1) Adversarial Robustness: Protecting against sophisticated attacks targeting agent decision-making processes, including prompt injection and KPI manipulation. (2) Decision Explainability and Audit-ability: Developing interpretable reasoning chains and audit trails to facilitate troubleshooting and regulatory compliance. (3) Fallback Mechanisms and Self-Diagnosis: Implementing robust failure recovery protocols and self-monitoring capabilities to maintain service continuity.

\subsubsection{Advanced Reasoning and Integrated Knowledge Bases}

Enhancing the cognitive capabilities of agentic systems requires fundamental advances in reasoning and knowledge management: (1) Wireless-Specific RAG Systems: Integrating 3GPP standards, research literature, and operational knowledge to ground decisions in domain expertise. (2) Long-Term Memory Architectures: Developing persistent knowledge systems that enable learning from extended operational history and cumulative experience. (3) Causal Reasoning Capabilities: Moving beyond correlation to understand causal relationships in network behavior for more fundamental problem-solving.

\subsubsection{Standardization and Ecosystem Interoperability}

Successful industry adoption depends on establishing open frameworks and collaborative ecosystems: (1) Open Interfaces and APIs: Standardizing agent-network and agent-agent interfaces to enable multi-vendor interoperability and innovation. (2) Benchmark Datasets and Scenarios: Developing comprehensive testing environments and performance metrics to evaluate and compare agentic approaches. (3) Industry Collaboration Frameworks: Establishing working groups within standards bodies to define architectural requirements, safety standards, and deployment guidelines.

\subsubsection{Two-Timescale Learning for Large-Scale Deployment}

Efficient adaptation across large-scale, heterogeneous 6G environments necessitates a two-timescale learning paradigm. (1) Short-timescale Memory-driven Adaptation: This enables rapid deployment through retrieval and reuse of validated experiential trajectories, thereby circumventing redundant exploration and cold-start inefficiencies. (2) Long-timescale Federated Learning (FL): By securely aggregating anonymized reflection diagnostics and optimization policies from multiple network domains, FL continuously refines the underlying agent priors and domain knowledge bases. This two-timescale architecture ensures that real-time control remains lightweight and deterministic, while background model evolution continuously enhances the reflection engine’s diagnostic and planning capabilities.

\section{Conclusion}
This article establishes simulation-in-the-loop validation for truly autonomous 6G networks and presents the first reflection-driven self-optimization framework that transforms agentic AI from an open-loop planner into a self-correcting system through continuous empirical validation and iterative refinement. Our closed-loop architecture orchestrates four specialized agents to enable autonomous resource management without human intervention. Extensive experiments demonstrate significant improvements over non-agentic approaches in terms of system throughput via escaping local optima, user QoS satisfaction through implicit intent recognition, and reduced resource utilization during low-traffic periods. This validates our core findings that reflection via simulation-in-the-loop is essential for agentic autonomous RAN for 6G.

\bibliographystyle{IEEEtran}
\bibliography{cite}

\clearpage
\onecolumn

\begin{figure}[!t]
\centering
\textbf{Appendix: Additional Large-Scale Experimental Results}

\vspace{2mm}

\subfloat[Reflection-enhanced optimization]{
    \includegraphics[width=0.83\linewidth]{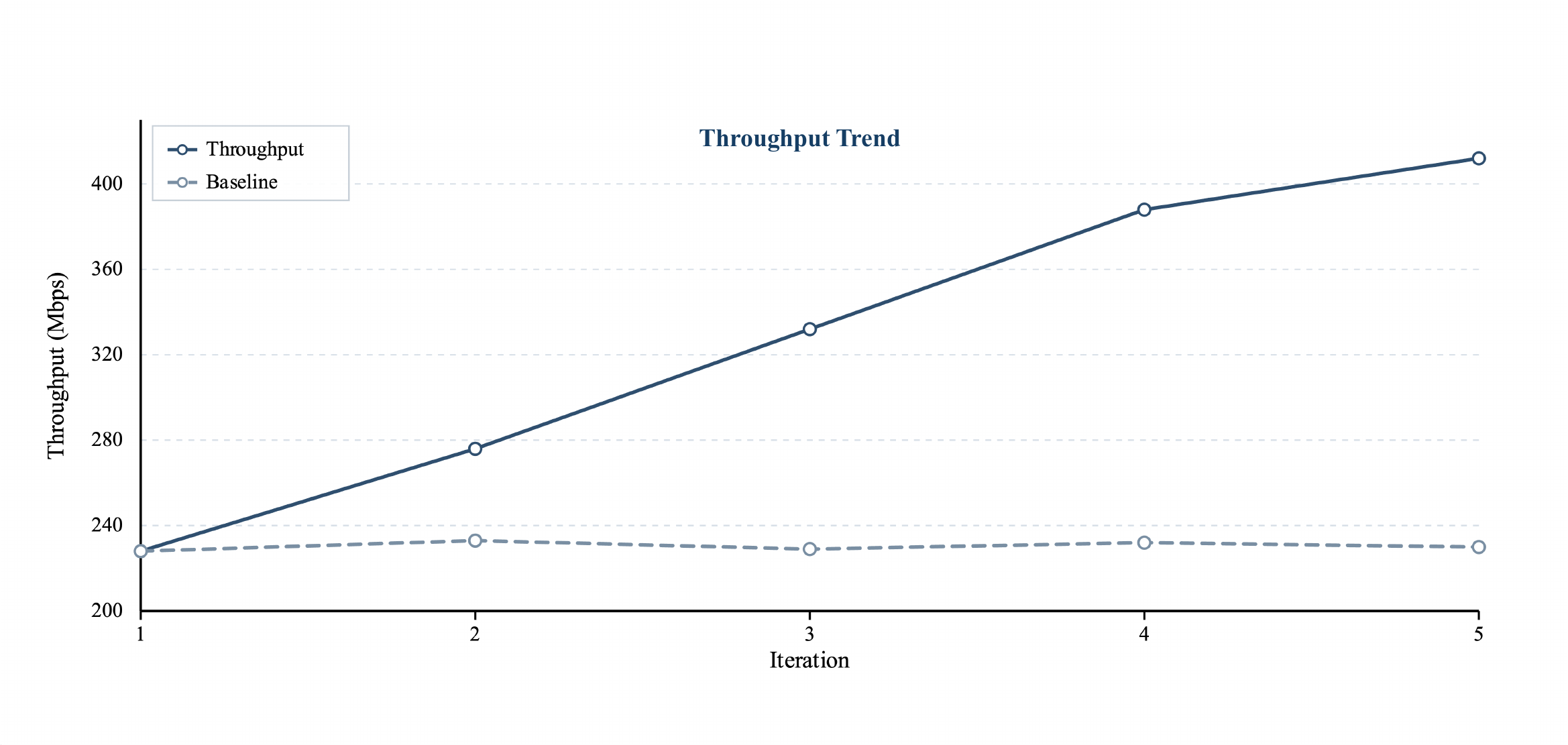}
    \label{fig:appendix_large_scale_a}
}

\vspace{1mm}

\subfloat[Reflection-enabled intent recognition]{
    \includegraphics[width=0.83\linewidth]{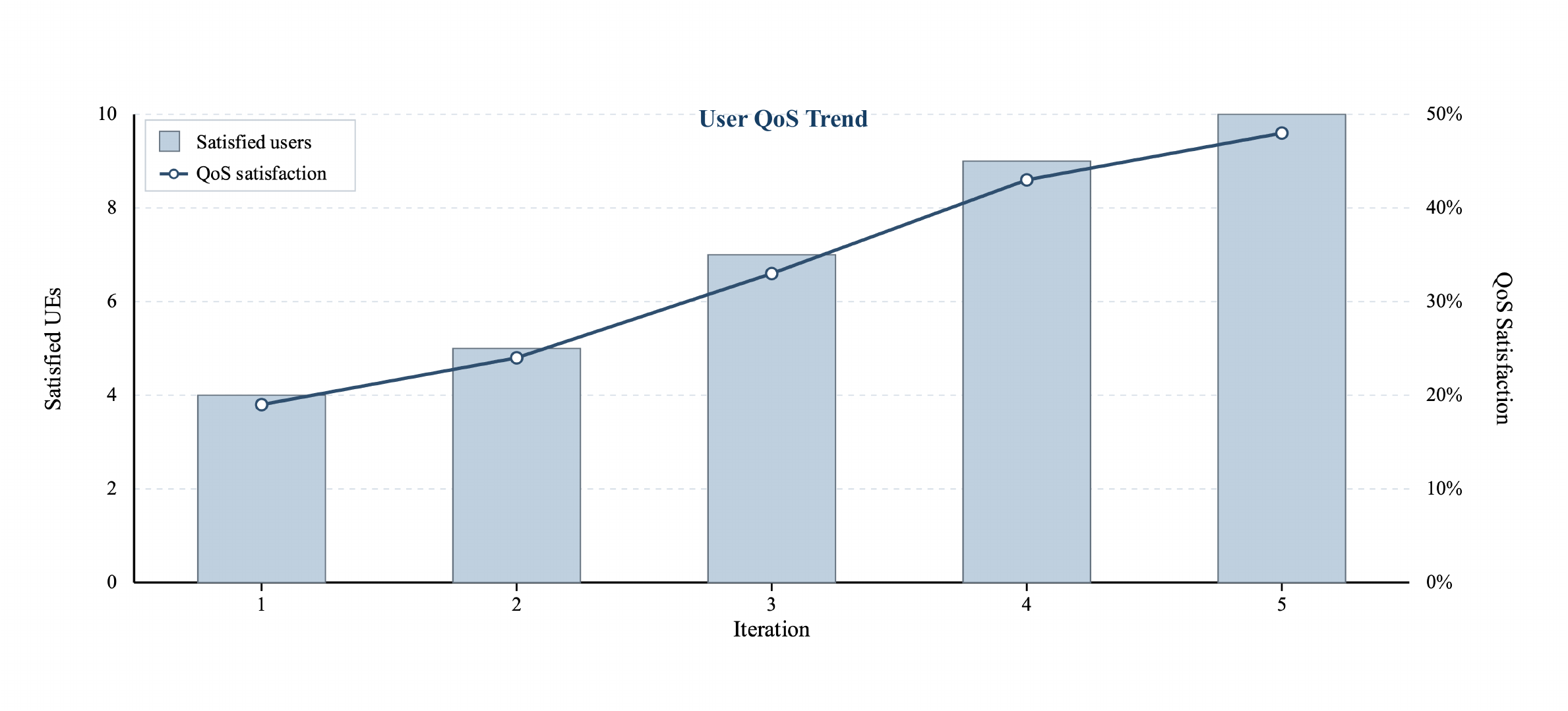}
    \label{fig:appendix_large_scale_b}
}

\vspace{1mm}

\subfloat[Reflection-driven context awareness]{
    \includegraphics[width=0.83\linewidth]{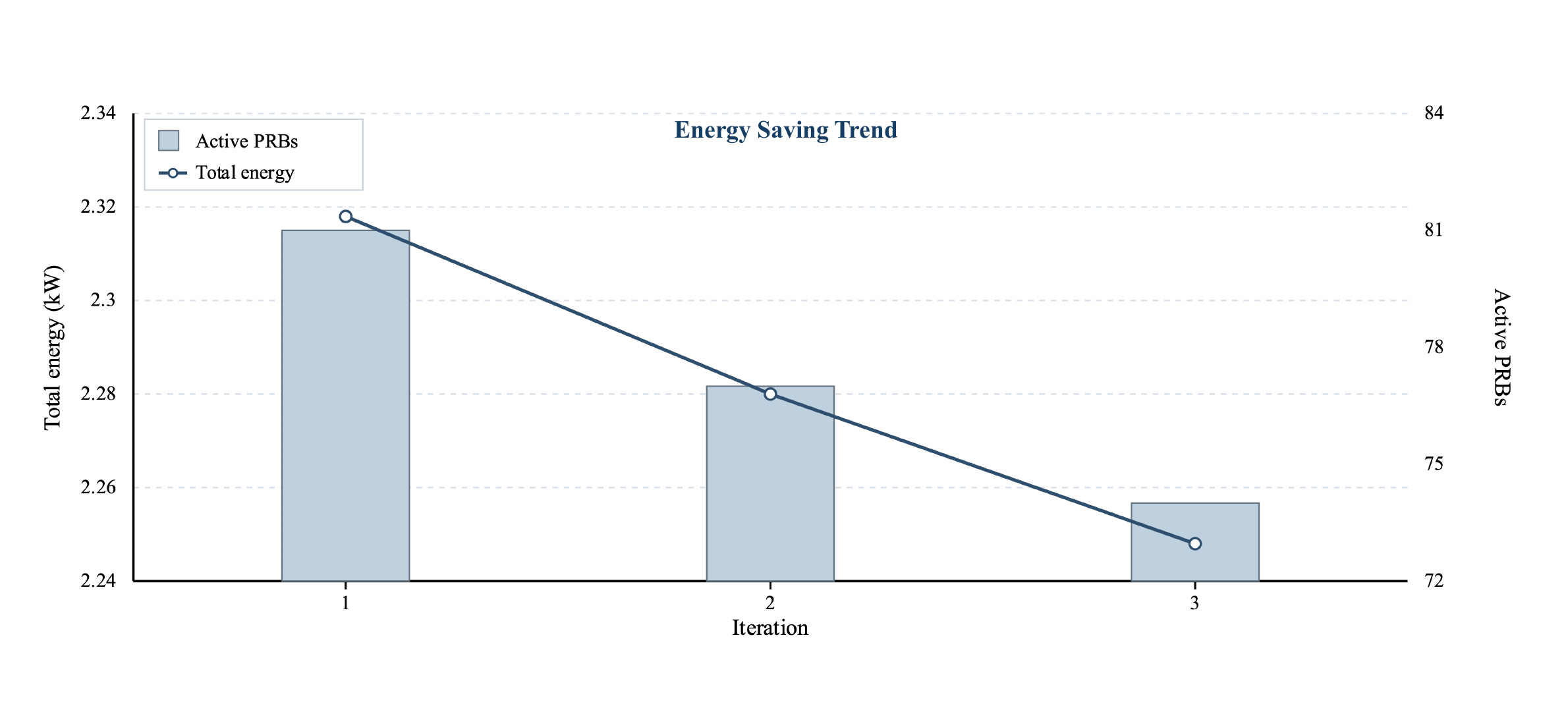}
    \label{fig:appendix_large_scale_c}
}

\caption{Additional large-scale results under a 7-cell, 21-UE, 84-PRB setting.}
\label{fig:appendix_large_scale}
\end{figure}

\end{document}